# Active Magnetoelectric Control of Terahertz Spin Current

Avinash Chaurasiya[1,⊥], Ziqi Li[2,⊥], Rohit Medwal[1]*, Surbhi Gupta[1], John Rex Mohan[3], Yasuhiro Fukuma[3,4], Hironori Asada[5], Elbert E.M. Chia[2]* & Rajdeep Singh Rawat[1]*

[1]Natural Sciences and Science Education, National Institute of Education, Nanyang Technological University, Singapore 637616, Singapore

[2]Division of Physics and Applied Physics, School of Physical and Mathematical Sciences, Nanyang Technological University, 21 Nanyang Link, Singapore 637371, Singapore

[3]Department of Physics and Information Technology, Kyushu Institute of Technology, Iizuka 820-8502, Japan

[4]Research Center for Neuromorphic AI Hardware, Kyushu Institute of Technology, Kitakyushu 808-0196, Japan

[5]Graduate School of Sciences and Technology for Innovation, Yamaguchi University, Ube 755-8611, Japan

[⊥]Co-first authors
*Corresponding authors

**Abstract:**

**Electrical control of photogenerated THz spin current pulses from a spintronic emitter has been at the forefront for the development of scalable, cost-efficient, wideband opto-spintronics devices. Artificially combined ferroelectric and ferromagnet heterostructure provides the potential avenue to control the spin dynamics efficiently utilizing the magnetoelectric coupling. The demonstration of the electric field control of spin dynamics has so far been limited up to gigahertz frequencies. Here, we demonstrate the electric field mediated piezoelectric strain control of photogenerated THz spin current pulse from a multiferroic spintronic emitter. The phase reversal of the THz spin current pulse is obtained from the combined effect of piezoelectric strain and a small constant magnetic field applied opposite to the initial magnetization of the ferromagnet. The piezoelectric strain-controlled phase switching of THz spin current thus opens a door to develop efficient strain engineered scalable on-chip THz spintronics devices.**

**Introduction:**

Artificial multiferroic systems, consisting of ferroelectric (FE) and ferromagnetic (FM) orders[1-6], provide a unique approach to control the ferromagnetic order by electric field or the ferroelectric order using magnetic field[7,8] due to their magnetoelectric coupling[8,9]. Utilizing magnetoelectric coupling in multiferroic FE/FM heterostructures, the active manipulation of various processes has been demonstrated[8,10-15] but the active manipulation of ultrafast terahertz (THz) spin dynamics and associative superdiffusive processes remain unexplored. The recent progress in THz spintronics suggests a typical ferromagnet (FM)/heavy metal (HM) heterostructure upon illumination by a femtosecond (fs) laser pulse drives the spins out of equilibrium[16] in the FM layer causing the ultrafast demagnetization[17-19] and ballistic superdiffusion of spins[20] into the adjacent HM layer[20,21] generating a transient charge current through the spin to charge conversion process of inverse spin Hall effect (ISHE)[22,23]. This transient charged current, which exists in the sub-picosecond (ps) timescale, emits THz electromagnetic radiation according to Maxwell's equation[17,24,25]. Achieving the active control of such sub-picosecond THz spin current pulses and the resulting THz radiation pulse emitted from the fs photoexcitation of THz spintronic emitter is highly desirable to fabricate the on-chip opto-spintronic devices for potential applications in high-speed computing[26] and memory devices with different functionalities[27]. The control of the THz spin current through electrical current[28], magnetic field[29,30], photothermal effect[31], anisotropy modulation[32] and helicity of light[29], have been demonstrated. However, the deterministic control of THz spin current through piezoelectric strain, using electric field, has not been explored to date.

In this article, we design an artificial multiferroic spintronic emitter (Cu/Pb(Mg$_{1/3}$Nb$_{2/3}$)O$_3$–0.31PbTiO$_3$(PMN-PT)/NiFe/Pt) to demonstrate the piezoelectric strain control of the THz spin current pulse amplitude as well as the phase switching. The sequential cyclic sweep of applied electric field ($E$) and corresponding measurements of strain, effective magnetization ($M_{eff}$) and emitted THz radiation pulse amplitude exhibit the remarkable correlated nonlinear butterfly hysteresis in *Strain-E*, $M_{eff}$-*E* and *THz-E* curves indicating simultaneous control and probing of strain dynamics, effective magnetization, and THz spin dynamics. The observed inversion of both the '*THz-E* butterfly hysteresis' and the magnetization vs magnetic field (*M H*) hysteresis curves upon changing the measurements configuration from easy-axis to the hard-axis of magnetization, clearly demonstrates that the manipulation of the emitted THz radiation pulse is due to the piezoelectric strain control of the easy/hard magnetization axis of NiFe layer. The phase switching of the emitted THz radiation pulse is demonstrated using the combined effect of electric field induced strain and a small constant magnetic field applied opposite to the initial magnetization direction of the NiFe layer. This phase switching of THz spin current using piezoelectric strain in artificial multiferroic heterostructures is a milestone step forward to write the information in the form of bit "0" and "1" for development of highly desirable all electric field controlled strain engineered on-chip THz spintronics devices.

**Results and discussion:**

To demonstrate the electric field control of THz spin current in an artificial multiferroic spintronic emitter via strain mediated manipulation of the effective magnetization, a hetero-structured stack of electrode/FE/FM/HM (Cu/PMN-PT/NiFe/Pt) has been designed (refer the "Methods" section for fabrication details), as shown in Figure 1(a). The externally applied electric field across the device stack induces strain that cause respective compression and elongation along x- and y-directions ascribed to negative ($d_{31} < 0$) and positive ($d_{32} > 0$) piezoelectric charge coefficients along these directions. Figure 1(a) also shows the schematics of the unit cells of rhombohedral (0 1 1) PMN-PT, poled perpendicular to (0 1 1) plane, and their corresponding maximum strained state at $+8 \text{ kV cm}^{-1}$ along [0 1 1] with net out-of-plane polarization, solid red arrow, and minimum strained state at coercive electric field, $E_c$, of $-2 \text{ kV cm}^{-1}$ along [0 $\bar{1}$ $\bar{1}$] with zero net polarization. These electric field values, required to attain maximum and minimum strained states, are estimated from in-situ X-ray diffraction (XRD) measurements of the PMN-PT. The XRD spectra at $+8 \text{ kV cm}^{-1}$ and coercive electric field of $-2 \text{ kV cm}^{-1}$ in Figure 1(b), with maximum shift in the (0 1 1) diffraction peak, correspond to the maximum and minimum strained states of the device, respectively. Further, to obtain the variation of the strain, calculated from the shift in (0 1 1) $K_{\alpha_1}$ diffraction peak, the XRD measurements were performed with the cyclic sweep of electric field as shown in Figure 1(c). The XRD patterns obtained at various electric field applied in cyclic order, from $+8$ kV cm$^{-1}$ → $-8$ kV cm$^{-1}$ → $+8$ kV cm$^{-1}$, are shown in supplementary section S1. Figure 1(c) show a butterfly loop like behavior (*Strain-E* hysteresis) wherein maximum-strained (at $+8$ kV cm$^{-1}$) and minimum-strained (at coercive field of $-2$ kV cm$^{-1}$) states can be identified for electric field sweep from $+8$ kV cm$^{-1}$ to $-8$ kV cm$^{-1}$.

The induced strain is utilized to control the effective magnetization ($M_{eff}$) of NiFe layer through magnetoelectric coupling between PMN-PT and NiFe. The change in $M_{eff}$ of NiFe layer is investigated using ferromagnetic resonance (FMR) spectroscopy. Figure 1(d) shows the shift in the FMR peak position for maximum and minimum strained states confirming the modulation of $M_{eff}$ of NiFe layer by an applied electric field induced strain. The FMR spectra for various strained states, for the cyclic variation of applied electric field from $+8$ kV cm$^{-1}$ → $-8$ kV cm$^{-1}$ → $+8$ kV cm$^{-1}$, are given in supplementary section S2. It is observed that the cyclic variation of the $M_{eff}$ obtained from analyzing resonance spectra with the applied electric field also follows a butterfly behavior ($M_{eff}$-*E* hysteresis), see Figure 1 (e), similar to the *Strain-E* hysteresis. The maximum and minimum values of the $M_{eff}$ are observed at $\pm 8$ kV cm$^{-1}$ and $\mp 2$ kV cm$^{-1}$ respectively, which is in direct correlation with the strain in the device. Therefore, the applied electric field can simultaneously control the strain and $M_{eff}$ in the device.

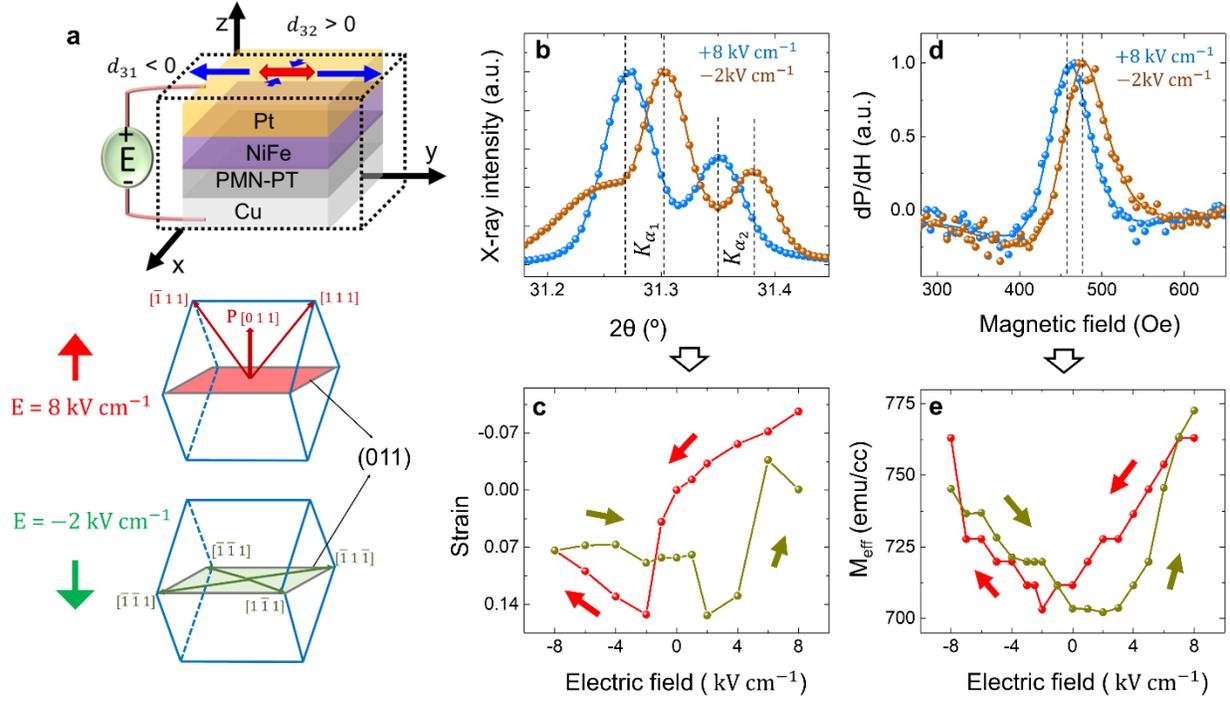

Figure 1: **Electric field control of strain and effective magnetization.** (a) Schematic of the artificial multiferroic heterostructured spintronic device stack with the applied electric field. The middle and bottom schematics show the rhombohedral unit cells of the PMN-PT at maximum ($+8$ kV cm$^{-1}$) and minimum strained ($-2$ kV cm$^{-1}$) states. (b) X-ray diffraction patterns about (011) diffraction peak at maximum and minimum strained states in the PMN-PT (011) single crystal. (c) *Strain-E* butterfly hysteresis for a PMN-PT (011) substrate. (d) Ferromagnetic resonance spectra recorded at maximum and minimum strained states. (e) Variation of the effective magnetization, $M_{eff}$, with the applied electric field ($M_{eff}$-*E* butterfly hysteresis).

After the confirmation of the electric field induced strain control of $M_{eff}$, the piezoelectric strain manipulation of the magnitude and phase of the emitted THz pulses is probed using THz time domain spectroscopy setup shown in Figure 2(a) (refer the "Methods" section for setup). Figure 2(b) shows the schematic of the device with mechanism of THz radiation emission. Here, the photoexcitation leads to ultrafast demagnetization and ballistic superdiffusion of spins in NiFe layer at sub-ps time scale which results in the injection of THz spin current ($\boldsymbol{j_s}$) pulse from the FM to the HM Pt layer where the $\boldsymbol{j_s}$ is converted into a charge current ($\boldsymbol{j_c}$) pulse through the ISHE, according to $\boldsymbol{j_c} = \gamma \boldsymbol{j_s} \times \boldsymbol{M}/|\boldsymbol{M}|$[28] with $\gamma$ as spin Hall angle and $M$ being the magnetization of the ferromagnet. The transient THz charge current in HM results in the emission of THz radiation which is recorded to probe the characteristics of photogenerated THz spin currents at different applied electric field. Figure 2(c) shows the emitted THz radiation pulses at the maximum strained (+8 kV cm$^{-1}$) and unstrained (0 kV cm$^{-1}$) states, at remanent magnetization, for measurements along the magnetization hard-axis (H.A.) of NiFe. The identification of hard/easy axis of magnetization of NiFe layer is done by carefully analyzing the emitted THz pulse amplitudes for maximum strained and unstrained states at different orientations of the sample. The significant enhancement in the amplitude of the emitted THz pulse at maximum strained state (+8 kV cm$^{-1}$) in comparison to the zero electric field state, in Figure 2(c), is due to the enhancement in the remanent magnetization of the ferromagnetic layer which leads to the manipulation of the spin current $j_s$, ballistically diffusing in the HM layer, generating a charged current $j_c$ and hence changing the amplitude of the emitted THz radiation. Further, to verify the piezoelectric strain control of THz amplitude the emitted THz radiation pulses were recorded by sweeping the electric field in cyclic order from +8 kV cm$^{-1}$ → −8 kV cm$^{-1}$ → +8 kV cm$^{-1}$. The emitted THz radiation pulse signal recorded at different applied electric fields are presented in supplementary section S3. The plot of THz radiation pulse amplitude as a function of applied electric field, in Figure 2(d), exhibits a non-linear butterfly loop like curve which we have termed as '*THz-E* butterfly hysteresis'. The obtained '*THz-E* butterfly hysteresis', in Figure 2(d), is in direct correlation with the *Strain-E* and $M_{eff}$-*E* butterfly hysteresis shown in Figure 1(c) and 1(e) respectively. This demonstrates the simultaneous electric field control of the strain, the $M_{eff}$ and the THz pulse amplitude in multiferroic opto-spintronic device.

The change in the spin Hall angle of the Pt and the change in the excitation power of the laser pulse transmitting through the PMN-PT substrate during the cyclic sweep of electric field are ruled out as the possible causes of butterfly loop like curves, due to no change in the detected THz amplitude for measurements performed at in-plane magnetization saturation field of 800 Oe (discussed in Supplementary Information Section S4). The *THz-E* butterfly loop is not observed at applied magnetic field of 800 Oe (nearly constant THz radiation pulse amplitude is observed) indicating negligible electric field modulation of the ISHE.

Thereafter, the device is rotated to the easy-axis of magnetization of NiFe layer to investigate the THz radiation emission for different electric field at remanent magnetization. Figure 2(e) shows the THz pulse signals at maximum strained and untrained states of the device for this orientation. The THz pulse signals recorded for various other strained states are shown in supplementary section S4. The significant decrease in the THz pulse amplitude at maximum strained state in comparison to unstrained state is due to the decrease in the remanent magnetization which leads to injection of lower spin current $\boldsymbol{j_s}$ from FM to HM layer thus smaller $\boldsymbol{j_c}$. The variation in THz pulse amplitude with the electric field swept in cyclic order from $+8$ kV cm$^{-1}$ → $-8$ kV cm$^{-1}$ → $+8$ kV cm$^{-1}$ is plotted in Figure 2(f) which also exhibits butterfly loop like hysteresis. It is observed that the '*THz-E* butterfly hysteresis loop' in Figure 2(f) is inverted in comparison to the one shown in Figure 2(d).

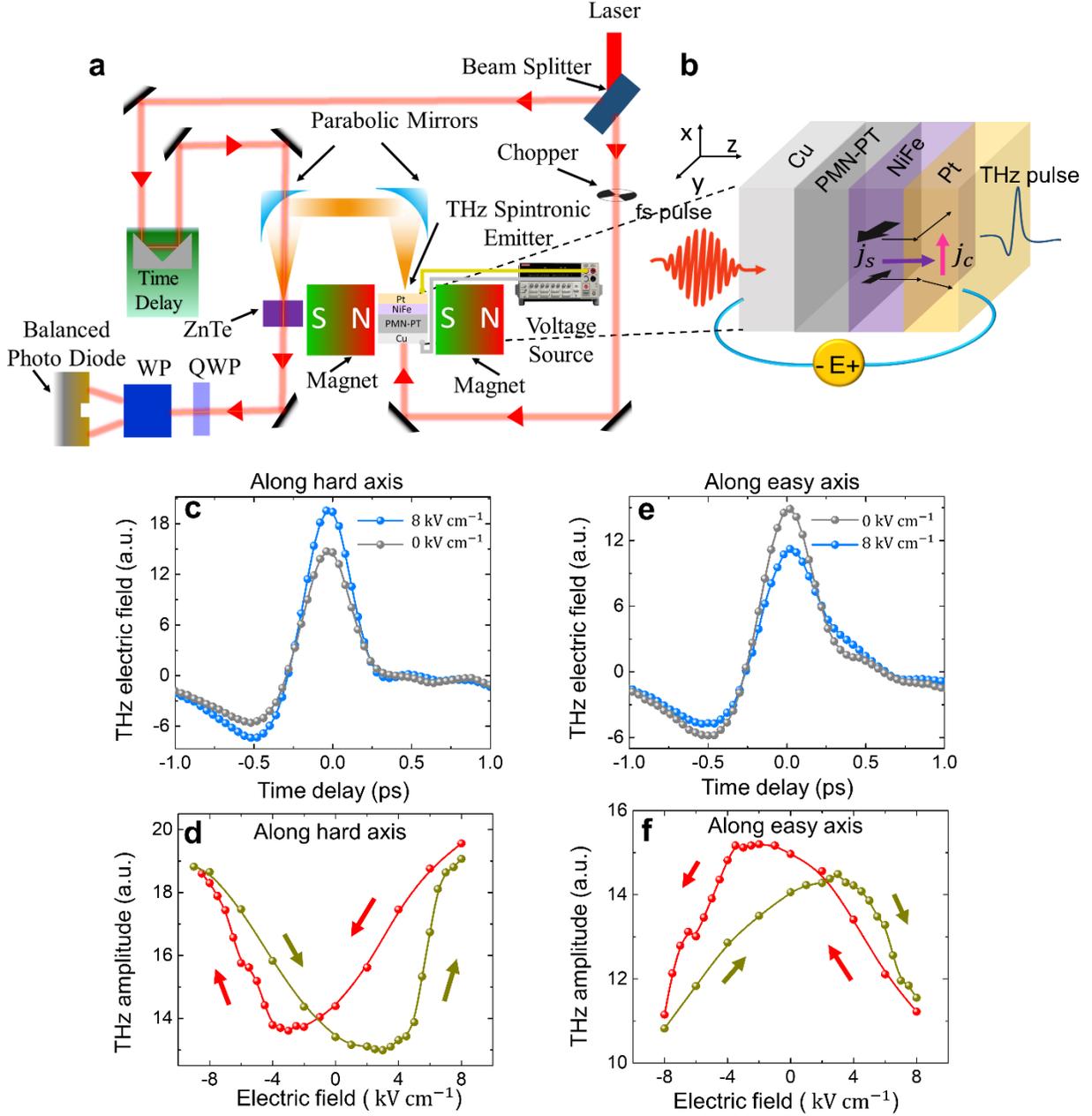

Figure 2: **Electric field control of THz spin current at remanent magnetization.** (a) Schematic of the setup for measurement and generation of THz radiation, at 1.2 mJ cm$^{-1}$ pump fluence. (b) Schematic of the artificial multiferroic spintronic emitter with the mechanism for emission of THz spin current. (c) Emitted THz spin current at applied electric field of 0 kV cm$^{-1}$ and +8 kV cm$^{-1}$, for measurements set along hard axis of magnetization of NiFe. (d) *THz-E* butterfly hysteresis along the hard axis of magnetization of NiFe. (e) Emitted THz spin current at applied electric field of 0 kV cm$^{-1}$ and +8 kV cm$^{-1}$, for measurements set along easy axis of NiFe. (f) *THz-E* butterfly hysteresis along the easy axis of magnetization of NiFe.

To understand the origin of the inversion of '*THz-E* butterfly hysteresis' (Figure 2(d) and 2(f)), *M-H* measurements were performed using VSM (vibration sample magnetometry). Firstly, the easy and the hard axis of magnetization of NiFe layer were identified using VSM measurement by in-plane rotation of the sample in step size of 10°. Figure 3(a) shows the variation of the normalized remanent magnetization with the complete in-plane rotation of the sample from 0° to 360° in the absence of electric field. Here, it is evident that the sample has the easy axis along 90° and 270° and the hard axis along 0° and 180°, as shown in Figure 3(b). After finding the easy and hard axis of magnetization of FM layer, the *M-H* hysteresis measurement were performed at different applied electric fields. Initially, we set the sample position along the hard axis of magnetization and performed the *M-H* hysteresis measurements at $0 \text{ kV cm}^{-1}$ and $+8 \text{ kV cm}^{-1}$, shown in Figure 3(c). It is observed that at $+8 \text{ kV cm}^{-1}$ the $M_r/M_s$ (normalized remanent magnetization, marked by dotted line in Figure 3(c)) of the *M-H* hysteresis loop increases in comparison to the hysteresis loop at $0 \text{ kV cm}^{-1}$, which is in direct correlation with the observed increase in the emitted THz radiation pulse amplitude observed in Figure 2(c). The increase in the squareness, $M_r/M_s$ ratio, of *M-H* hysteresis loop from unstrained to maximum strained state also indicates the rotation of hard axis of magnetization of NiFe. The sample is then rotated by 90° to align it along the easy axis of magnetization and corresponding *M-H* hysteresis measurements at the electric fields of $0 \text{ kV cm}^{-1}$ and $+8 \text{ kV cm}^{-1}$ are shown in Figure 3(d). A *reverse* trend, with a decrease in $M_r/M_s$ value at $+8 \text{ kV cm}^{-1}$ in comparison to the $0 \text{ kV cm}^{-1}$, is observed, which is consistent with the smaller THz amplitude observed in Figure 2(e) for this orientation, leading to the inversion of '*THz-E* butterfly hysteresis'. The decrease in the squareness of *M-H* hysteresis loop from unstrained to maximum strained state indicates the rotation of easy axis of magnetization of NiFe. This confirms that the THz spin current pulse amplitude is manipulated by the rotation of the magnetization easy/hard axis of ferromagnetic layer under piezoelectric strain modulated magnetic anisotropy in multiferroic THz spintronic emitter.

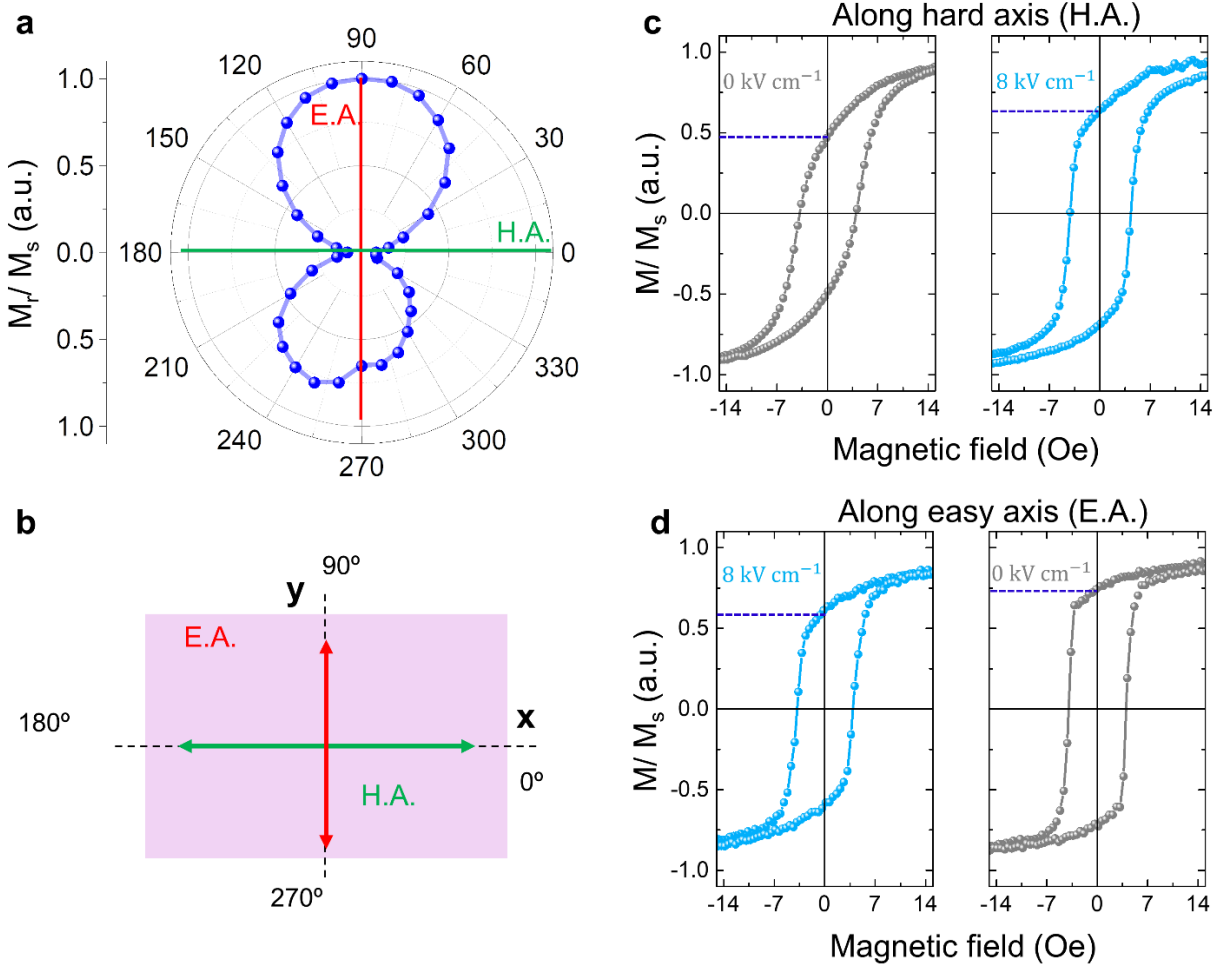

Figure 3: **Piezoelectric strain control of remanent magnetization.** (a) Variation of normalized remanent magnetization as a function of in-plane rotation of sample. (b) Schematic of the easy and hard axis of magnetization in a sample plane. (c) *M-H* hysteresis along hard axis of the magnetization at 0 kV cm$^{-1}$ and +8 kV cm$^{-1}$. (d) *M-H* hysteresis along easy axis of the magnetization at 0 kV cm$^{-1}$ and +8 kV cm$^{-1}$. The dotted horizontal line in (c) and (d) indicates the normalized remanent magnetization, $M_r/M_s$, which changes with the change in applied electric field.

It is well known that the magnetization in magneto-strictive FM materials cannot be switched by piezoelectric strain alone, for phase reversal of THz spin current pulse and hence a suitable combination of electric and magnetic field needs to be used. To deduce the suitable combination of electric and magnetic field values for THz pulse phase reversal, the *THz-H* hysteresis were plotted by recording the peak amplitude of the THz radiation pulse at different applied magnetic field values for zero and 10 kV cm$^{-1}$ electric fields, as shown in Figure 4(a). The whole set of *THz-H* hysteresis measurements for different applied electric field values is given in supplementary section S6. The measured *THz-H* hysteresis at different applied electric field values further confirms the control of THz amplitude at remanent magnetic field. In addition to the control over the THz pulse amplitude, the piezoelectric strain combined with the laser heating effect gives the control of the magnetic coercivity and *THz-H* hysteresis area as seen in Figure 4(a) and in supplementary section S6.

*THz-H* hysteresis loops at $+0$ kV cm$^{-1}$ and $10$ kV cm$^{-1}$ in Figure 4 (a) suggest that the phase of the THz spin current pulse can be switched at a selective small applied magnetic field opposite to the initial magnetization of FM. It is found that at magnetic field of 1.5 Oe, marked by blue arrow in Figure 4(a), the phase of the THz pulse can be reversed just by switching the electric field from 0 kV cm$^{-1}$ to 10 kV cm$^{-1}$, as shown in Figure 4(b). The applied magnetic field of 1.5 Oe is not unique, rather it exists in a range from 1.3 Oe to 2.3 Oe between the coercive magnetic field values of *THz-H* hysteresis for 0 kV cm$^{-1}$ and 10 kV cm$^{-1}$. It may be noted, however, that, the switching off the electric field from 10 kV cm$^{-1}$ to 0 kV cm$^{-1}$ does not reverse the phase back to the initial state. This is because the direction and magnitude of the magnetic field remains the same, 1.5 Oe, and the piezoelectric strain manipulates the $M_{eff}$ in the direction same as that of 1.5 Oe. The phase reversal of the emitted THz spin current can, however, be realized again if the direction of the applied magnetic field is reversed.

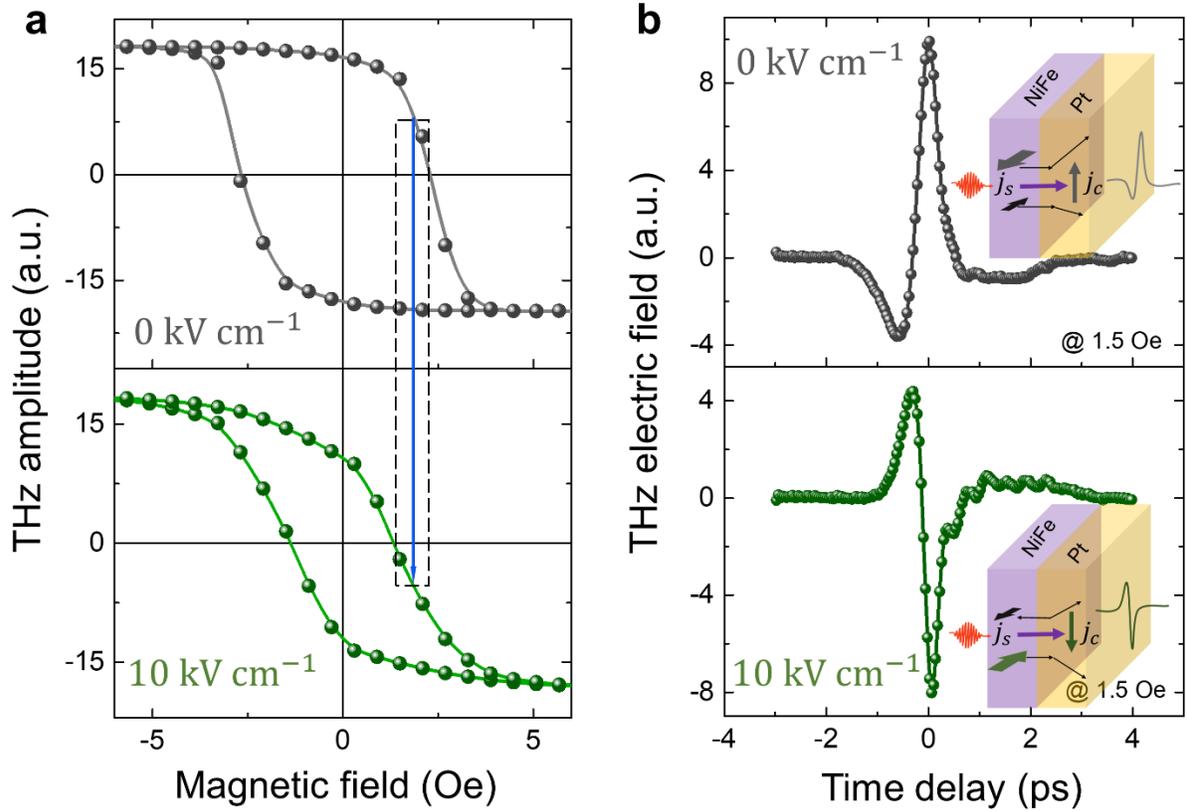

Figure 4: **Electric field controlled phase switching of THz spin current pulse.** (a) *THz-H* hysteresis loop at 0 kV cm$^{-1}$ and 10 kV cm$^{-1}$. The blue arrow marks the applied magnetic field value of +1.5 Oe, in the direction opposite to initial magnetization direction of FM layer where the THz pulse phase reversal should be observed, as shown in (b). The dotted box marks the region between the coercive field points depicting the magnetic field range, opposite to initial magnetization direction, with in which the phase reversal can be achieved. (b) Phase switching of the THz pulse with the change in electric field from 0 kV cm$^{-1}$ to 10 kV cm$^{-1}$ at small applied magnetic field of +1.5 Oe.

In summary, we demonstrate piezoelectric strain control of THz spin current pulse in a multiferroic spintronic emitter where the strain manipulates the injected spin current from FM to HM layer by controlling the effective magnetization due to the change in magnetic anisotropy of the FM layer. Moreover, the combined effect of strain and magnetic field switches the phase of the emitted THz pulse by reversal of the effective magnetization of the FM. Our results of strain engineered emitted THz radiation have far reaching implications for research on multiferroics, ultrafast magnetism and THz science. The experiments demonstrate a novel methodology to probe and control the spin and strain dynamics by correlating them with emitted THz radiation from the multiferroic spintronic emitter in ultrafast timescales. Piezoelectric strain-controlled photogenerated THz spin current thus opens a door to develop futuristic high speed opto-spintronic devices for computing and communication technologies.

**Methods**

**Device Fabrication:** A ferroelectric (011) PMN-PT substrate, with both side polished surface is used to fabricate the THz spintronic device. The NiFe (5 nm) and Pt (5 nm) thin films were deposited on PMN-PT, in that order, by electron beam at the base pressure of $10^{-6}$ Pa. The deposition rate of the NiFe was kept 0.025 Å/s while for Pt, the deposition rate was kept at 0.03 Å/s. The chamber pressure during the deposition was 1.68-1.84 $\times 10^{-5}$ Pa. For the capacitive configuration device fabrication, the Cu electrode was deposited on the bottom surface of the PMN-PT using 50 W DC magnetron sputtering at an argon pressure of 3 mTorr. The quality of the NiFe/Pt bilayer and NiFe thin film was characterized using the ferromagnetic resonance (see Section S10 in Supplementary Information). The value of effective magnetization and damping coefficient extracted from the measurement are found to be 820 mT and 0.022 demonstrating the good quality of the NiFe/Pt bilayer and NiFe thin film.

**Ferromagnetic Resonance (FMR) Measurements:** The FMR measurements are performed using Cryo-FMR. The electrical connections to the top (Pt) and bottom (Cu) electrodes were made using the Cu wire and the electric field was applied using Keithley 2410 source meter during FMR measurements.

**I-V Measurements:** The I-V characteristic of the device is performed using the Keithley 2410 source meter at room temperature as well as at different laser pump fluence. The laser heating effect on the coercive electric field was observed and given in supplementary section S7.

**X-ray diffraction:** High resolution out of plane X-ray diffraction measurement were performed on (011) PMN-PT substrate at different applied electric field using the Keithley 2410 source meter by a Bruker D8 Discover diffractometer (Cu K$_\alpha$ X-ray source) setup.

**THz Generation and Measurement Setup:** A Ti:sapphire ultrafast amplifier (Legend Coherent, 800 nm, ~50 fs, 1 KHz) was employed in this setup. The generated laser pulses are split into two optical paths by a pellicle beam splitter: One is used to pump at normal incidence and generate spin-polarized carriers and the other portion of the laser beam is used for THz detection. The THz pulses generated were projected onto the x-axis (vertical) by using a wire-grid THz polarizer, and then collected and focused by two parabolic mirrors onto a 1 mm thick <110> ZnTe crystal for free-space electro-optic sampling. The magnetic field direction is controlled by a motorized rotational stage (Thorlabs PRM1/MZ8). All measurements were performed at room temperature in dry air with a humidity less than 2%.

***THz-E* hysteresis measurement:** To observe the $THz\ E$ hysteresis curves, the emitted THz pulses were recorded at different applied electric field. The electric field is applied using the Keithley 2410 source meter. Variation of the THz pulse amplitude with the applied electric field gives the *THz-E* butterfly like hysteresis curve.

**Acknowledgement**

A.C. would like to acknowledge the NTU- Research Scholarship (NTU-RSS). R.M., S.G. and R.S.R. acknowledge the support from the Ministry of Education, Singapore grant no MOE2019-T2-1-058 (ARC 1/19 RSR) and National Research Foundation grant no NRF-CRP21-2018-0003. Z.L. and E.E.M.C. would like to acknowledge the support from the Singapore Ministry of Education AcRF Tier 3 Programme "Geometrical Quantum Materials" (Grant No. MOE2018-T3-1-002)". Y.F. would like to acknowledge the JSPS Grant-in-Aid (KAKENHI No. 18H01862a and 19K21112).